\documentstyle[12pt,aaspp4,flushrt]{article}

\def\eg.{{\it e.g.},}
\def\ie.{{\it i.e.}}
\def\cf.{{\it cf}}
\def\etal.{{\it et al.}}

\def\arcsec#1{\ifmmode{\if.#1{''\llap.}\else{''} #1\fi}\else
{\if.#1$''$\llap.\else$''$ #1\fi}\fi}
\def\kms.{km~s$^{-1}$}
\def\mic.{$\mu$m}
\def\um.{$\mu$m}
\def\msun.{$M_\odot$}
\def\lsun.{$L_\odot$}
\def\13CO.{$^{13}$CO}
\def\C18O.{C$^{18}$O}

\def\simless{\mathbin{\lower 3pt\hbox
  {$\rlap{\raise 5pt\hbox{$\char'074$}}\mathchar"7218$}}} 
\def\simgreat{\mathbin{\lower 3pt\hbox
  {$\rlap{\raise 5pt\hbox{$\char'076$}}\mathchar"7218$}}} 

\def\I.{\kern.2em{\small I}}  \def\II.{\kern.2em{\small II}} 
\def\III.{\kern.2em{\small III}} \def\IV.{\kern.2em{\small IV}}

\slugcomment{{\bf Revision Date :} 14 July 1999}

\lefthead{Herbst et al.}
\righthead{Deep Survey for Methane Brown Dwarfs}

\begin {document}

\title {Constraints on the Space Density of Methane Dwarfs
	and the Substellar Mass Function
	from a Deep Near-Infrared Survey}

\author {T. M. Herbst, D. Thompson\altaffilmark{1}, R. Fockenbrock,\\
	H.-W. Rix, and S. V. W. Beckwith\altaffilmark{2}}
\altaffiltext{1}{Now at Caltech, MS 320-47, Pasadena, CA 91125, USA}
\altaffiltext{2}{Now at STScI, 3700 San Martin Dr., Baltimore, MD 21218, USA}
\affil {Max--Planck Institut f\"ur Astronomie,}
\affil {K\"onigstuhl 17, 69117 Heidelberg, Germany}
\authoremail{herbst@mpia-hd.mpg.de}

\begin{abstract}

We report preliminary results of a deep near-infrared search for
methane-absorbing brown dwarfs; almost five years after the discovery
of Gl~229b, there are only a few confirmed examples of this type
of object.  New J~band, wide-field images, combined with pre-existing
R~band observations, allow efficient identification of candidates by
their extreme (R-J) colours. Follow-up measurements with custom
filters can then confirm objects with methane absorption.  To date, we
have surveyed a total of 11.4 square degrees to J$\sim$20.5 and
R$\sim$25. Follow-up CH$_4$ filter observations of promising
candidates in 1/4 of these fields have turned up {\it no} methane
absorbing brown dwarfs. With 90\% confidence, this implies that the
space density of objects similar to Gl~229b is less than 0.012 per
cubic parsec. These calculations account for the vertical structure of
the Galaxy, which can be important for sensitive measurements.
Combining published theoretical atmospheric models with our
observations sets an upper limit of $\alpha\leq$0.8 for the exponent of
the initial mass function power law in this domain.

\end{abstract}


\vskip 1.0 in

\section {Introduction}

Brown dwarfs are star-like objects that span the mass range between
the gas giant planets like Jupiter and the least massive
stars. Although luminous early in their evolution due to gravitational
contraction and deuterium burning, brown dwarfs have insufficient mass
to ignite and sustain hydrogen fusion in their cores. The discovery
and confirmation of a small number of brown dwarfs in the last few
years has renewed interest in these objects after more than a decade
of difficult and unsuccessful searches.

One object in particular, Gl~229b, has been the focus of widespread
study, since it is at least 10 times less luminous than the
lowest-mass stars, and its near infrared spectrum shows broad, deep,
methane absorption features such as those seen in Jupiter
(\markcite{Na_95}Nakajima \etal. 1995, \markcite{Op_95}Oppenheimer
\etal. 1995, \markcite{Gb_96}Geballe \etal. 1996).  Recently, several
groups have undertaken large surveys for older, field brown dwarfs
(\ie. using 2MASS, \markcite{Ki_99}Kirkpatrick \etal. 1999, and DENIS,
\markcite{Dl_97}Delfosse \etal. 1997). These programs have identified
significant numbers of more luminous substellar objects, but there are
only a handful of additional examples of methane absorbing, very low
luminosity brown dwarfs like Gl~229b, all reported very recently
(\markcite{St_99}Strauss \etal. 1999, \markcite{Bu_99}Burgasser
\etal. 1999).

In this paper, we report the preliminary results of a deep, wide-field
survey for such ``methane brown dwarfs.'' This survey employs a
two-step approach to this task: efficient identification of candidates
throught their red (R-J) colours, followed by confirmation through
methane filter imaging.

\section {Survey Strategy}

The deep H and K band methane absorption features are a distinctive
and defining feature of these objects.  Suitably designed custom
filters can provide enormous contrast between the methane absorption
and the adjacent continuum, a contrast inconsistent with any known
continuum emission process or non-methane absorption
feature. \markcite{Ro_96} Rosenthal et al. (1996) showed a detection
of Gl~229b using a (relatively inefficient) 1\% bandwidth variable
filter tuned to these wavelengths.

The bright sky background in the H and K photometric bands hampers
efficient, wide-field surveys. Instead, we take advantage of another
identifying property of brown dwarfs to pre-select candidates.  The
spectral energy distribution of old brown dwarfs peaks in the J~band,
(\markcite{Na_95}Nakajima \etal. 1995), and, coupled with molecular
absorptions at shorter wavelengths, this profile produces extreme
$(R-J)$ colours. For example, Gl~229b has $(R-J)\simgreat$9
(\markcite{Ma_96}Matthews \etal. 1996, \markcite{Go_98}Golimowski
\etal. 1998), much redder than even the coolest stars (an M6 dwarf has
$(R-J)\sim$4 - \markcite{Be_91}Bessel, 1991).

Efficient J band measurements are possible using wide-field, prime
focus cameras, and we further streamline the survey by using {\it
pre-existing} R~band observations from a high-redshift supernova
search (see section~3.1). Subsequent measurements of promising targets
with the custom filters then isolate the methane absorbing brown
dwarfs.  Such follow-up is essential in eliminating false detections
due to supernovae, asteroids, and high proper-motion objects, (some of
which are interesting in themselves). Extremely red galaxies can also
appear; such galaxies will be the subject of a separate follow-up
investigation.

The pre-existing R data have a typical (3$\sigma$) limiting magnitude
of R$\sim$25. We chose a limiting magnitude of J$\sim$20.5 for the
near infrared survey, in order to identify effectively all candidates
with (R-J)$\simgreat$5. Note that Gl~229b would have J$\sim$20 at a
distance of 100~pc, so this R-J survey is well-matched to the vertical
scale height of the Galactic disk.

\section{Observations and Results}

\subsection {R and J Band Surveys}

The R~band observations took place over the period 1995--1997 as part
of the High-Z Supernova Search Team cosmology project
(\markcite{Sch_98}Schmidt \etal. 1998).  This supernova survey
comprises over 300 high galactic latitude fields, each 15~arcminutes
square, located within a few degrees of the celestial equator. We
registered and averaged typically six individual flat-fielded
exposures to create the final R~frame. Photometric calibration came
from galaxy number counts and the magnitude relations in
\markcite{Me_91}Metcalfe \etal. (1991), a technique sufficiently
accurate ($\pm$0.2~mag) to identify candidates with extreme (R-J)
colours.

We observed 40 of these fields (2.6 square degrees in total) in the
J~band between 11 and 13 October 1997, using the Omega Prime near
infrared camera (\markcite{Bz_98}Bizenberger \etal. 1998) mounted at
the prime focus of the 3.5~m telescope on Calar Alto. J~band
observations of a further 140 fields (9.0 square degrees) took place
in April~1998, October~1998, and May~1999.  None of the
recently-reported methane dwarfs lie in our fields.  Weather
conditions were excellent for all measurements. We applied standard
sky subtraction and flat--fielding routines to remove the effects of
sky background and pixel to pixel gain variations. Several of the
UKIRT faint standards (\markcite{CH_92}Casali and Hawarden, 1992)
served as photometric reference. The final J~band mosaic images are
15\arcmin$\times$15\arcmin, well-matched to the R~fields.

We used the SExtractor package (\markcite{BA_96}Bertin and Arnouts
1996) to identify objects in the J~band mosaic and derive brightness
and morphology information. Using the J~band locations for the R~band
photometry produced a catalog of all objects detected in the
near-infrared. This approach focuses on potentially interesting
targets (\ie. large $R-J$) while avoiding the 80\% of R~detections which
are not seen at J.  Nevertheless, the 40 data sets from October 1997
yielded information on approximately 35,000 objects.

\subsection {Follow-Up Observations of Candidates}

A small number of sources, typically 1 per field, display very red
colours and merit further observations with the methane
filters. Figure~2 shows R and J~band sub-images of a candidate with
(R-J)$>$7.1. Between 11 and 16 October 1998, we observed a total of 44
such objects from the first set of 40~fields, using the K~band methane
filters in Omega~Prime. Again, standard sky-subtraction and
flat-fielding techniques removed the background and pixel to pixel
gain variations.  The (R-J) cut-off for follow-up varied by
approximately 0.1 mag from field to field, but in all cases was less
than (R-J)=5.6.

For convenience, we will hereafter refer to the methane filters as
$K_C$ (K~continuum: 1.95-2.2~$\mu$m), and $K_A$ (K~absorption:
2.15-2.4~$\mu$m).  $K_C$ and $K_A$ are not standard photometric
filters, but they correspond closely to the lower and upper halves of
the K~band, respectively. Figure~2 plots $(K_C-K_A)$ for the 44
candidates against the ``stellarity index'' determined by
SExtractor. A negative $(K_C-K_A)$ points to the presence of methane,
and a higher stellarity means more star-like. Also plotted is the
methane colour and stellarity of Gl~229b derived from the images shown
in Figure~1. None of the 44 candidates, and in particular, none of the
more star-like objects, remotely approach Gl~229b in terms of
$K_C$-$K_A$ colour. In fact, almost all are somewhat red, not
``blue,'' even over this very short range of wavelengths.

\section {The Space Density of Methane Brown Dwarfs like Gl~229b}

What is the upper limit to the space density of objects like Gl~229b
that is consistent with our seeing none?  The answer depends on both
counting statistics and the effective survey volume. Observing zero
events in a single counting experiment can occur if the average number
of events is nonzero.  To set a 90\% confidence level on the minimum
number of expected events, we must calculate the expectation value of
a Poisson distribution whose probability of zero events is 10\%. This
value is approximately 2.3. Hence, with 90\% confidence, our
observations set an upper limit of 2.3 to the mean number of methane
brown dwarfs in our survey volume.

The effective survey volume $V_{eff}$ depends in turn on the intrinsic
luminosity of the sources, the angular size and galactic coordinates
of the fields, and the vertical structure of the galaxy. The volume
enclosed by a field of solid angle $\Omega_f$ to a distance $r_{max}$
is ${1\over 3}\,\Omega_f\,r_{max}^3$. The forty fields then enclose a
total search volume of 350~pc$^3$. Depending on the vertical
distribution of the targets in the Galaxy, however, the combination of
high galactic latitudes and relatively large $r_{max}$ will reduce the
{\it effective} survey volume, ($V_{eff}$ refers to the equivalent
volume in the midplane). For a single field $i$ at galactic latitude
$b$, the effective volume is:

$$V^i_{eff}=\Omega_f\int_0^{r_{max}} r^2\,e^{-\vert
Z_0 + r\,\sin{}b\vert\over h_z} \,dr,\eqno{(1)}$$

\noindent
where $h_z$ is the scale height for the target objects and $Z_0=12$~pc
is the vertical displacement of the Sun with respect to the Galactic
midplane (\markcite{Gi_89}Gilmore, 1989). Due to kinematic heating of
the Galactic disk, $h_z$ is a function of the age of the
population. Gl~229b has a best-fit age between 2-4~Gyr based on
atmospheric models and its measured luminosity (\markcite{Al_96}Allard
\etal. 1996, \markcite{Bu_97}Burrows \etal. 1997,
\markcite{Ma_96}Matthews \etal. 1996). This corresponds roughly to the
mean age of F5V stars, which have $h_z$=190pc, (\markcite{AQ_76}Allen,
1976). Combining these effects, we calculate an effective survey
volume $V_{eff}=\Sigma V_{eff}^{i}\sim$190~pc$^3$, and we can say with
90\% confidence that the space density for objects like Gl~229b is
less than $n_0=2.3~/~190=0.012$~pc$^{-3}$. For a scale height
appropriate to the minimum possible age for Gl~229b (0.5 Gyr,
\markcite{Al_96}Allard \etal. 1996), our $V_{eff}$ drops to 165~pc$^3$
and $n_0 < 0.014$.  For faint objects like Gl~229b, our type of deep,
relatively small area search is very effective.  In fact, in a single
night of observations, we cover a faint-object survey volume
comparable to the entire DENIS project.

\section {Limits on the Mass Function of Methane Brown Dwarfs}

The upper limit to the space density established in the previous
section refers to brown dwarfs ``similar to Gl~229b,'' that is,
extremely red objects with $M_J\approx 15.4$ and deep methane
absorption in the K~band. Younger or more massive objects will be
brighter and therefore visible to a greater distance. On the other
hand, brown dwarfs spend a relatively small fraction of their
lifetimes in this early, hot phase, and higher effective temperatures
may not allow the formation of CH$_4$.  In this section, we combine
published theoretical atmospheric models with our observations to
estimate the number of methane absorbing substellar objects that we
would expect in our survey. The probability of seeing such an object
depends on its mass, age, and the volume within which we could detect
it:

$$ d\langle N\rangle =
P_m(m)\,P_t(t)\,P_V(m,t)\,dm\,dt\,dV\eqno{(2)}.$$

$P_m$ is the substellar initial mass function (IMF), with suitable
normalization to give the number of objects per cubic parsec.  We
adopt the standard power law form of the IMF: $dn(m) = C\cdot
m^{-\alpha}\,dm$. Here, $dn(m)$ denotes the number of objects per
cubic parsec with mass between $m$ and $m+dm$ and $C$ is a
constant. Surveys of M~dwarfs within 8~pc of the sun give a local
space density of $\sim$0.065~pc$^{-3}$ and a power law exponent
$\sim$0.8 (\eg. \markcite{Le_97}Leinert \etal. 1997,
\markcite{HM_92}Henry and McCarthy 1992). Setting the integral of the
mass function for M~dwarfs to this space density gives the IMF
normalization at the M~Dwarf - Brown~dwarf boundary. We do not assume
that the exponent has the same value for brown dwarfs. Continuity of
the mass function at the boundary then requires that $C$ be a function
of $\alpha$.

The second term in equation~2, $P_t(t)$, is the age distribution of
substellar objects. For simplicity, we assume a constant mean galactic
star formation rate, $P_t = t_{max}^{-1}$, where $t_{max}$ is
$\sim$9~Gyr, the maximum age of the Galactic disk population
(\markcite{Wi_87}Winget \etal., 1987). We adopt a lower bound to the
age of field brown dwarfs, $t_{min}=0.5$ Gyr, typical of the minimum
age of stars in the solar neighbourhood and the maximum age of stars
in identifiable clusters (also see Section~\ref{discussion}).

Although young substellar objects are bright, successful
classification in our survey requires the presence of detectable
methane.  Strong $CH_4$ absorption will not occur above a certain
critical effective temperature, $T_c\approx 1200$~K
(\markcite{Ts_95}Tsuji \etal. 1995, \markcite{BS_99}Burrows and Sharp
1999). Because the effective temperature $T_{eff}$ is a function of
the age (and mass) of the object, $P_t(t)$ includes a factor
$f_{CH_4}$ describing the detectability of methane. We use an analytic
interpolation of the $T_{eff}$ curves of \markcite{Bu_97}Burrows
\etal. (1997) to calculate the effective temperature for each mass and
age, and then set $f_{CH_4}$ to zero or one depending on whether
$T_{eff}>T_c$ or vice versa.

$P_V(m,t)$ is just $V_{eff}$ from equation~1, with the upper
integration limit corresponding to the current values of $m$ and
$t$. \markcite{Bu_97}Burrows \etal. (1997) calculate the $M_J$ for
substellar objects as a function of mass and age, and again we employ
an analytic interpolation of their curves.  The scale height $h_z$
depends on the age $t$ of the object and is calculated based on a
smooth fit to the data in section~119 of \markcite{AQ_76}Allen (1976).

Combining these elements leads to a numerically solvable equation for
the expected number of substellar objects:

$$\langle N(\alpha)\rangle = \int_{0.02}^{0.075 M_\odot}
dm\,C\,m^{-\alpha} \,\int_{t_{min}}^{t_{max}}dt\,{f_{CH_4}\over
t_{max}}\,\sum_{{\em fields}\,i}\,
d\Omega\,\int_0^{r_{max}}dr\,r^2\,e^{-\vert
Z_0 + r\,\sin{}b\vert\over h_z},\eqno{(3)}$$

\noindent

\noindent
Solving Equation~3 with $\langle N\rangle=2.3$, the upper limit to the
number of detections in our survey, gives a limit on $\alpha$, the
power law index of the mass function. With 90\% confidence, our
observations demonstrate that $\alpha\leq$0.8 in the substellar
regime. Note that this conclusion allows continuity of the power
law index across the hydrogen burning limit.

\section{Discussion}
\label {discussion}

\markcite{Ki_99}Kirkpatrick \etal. (1999) report the discovery of 7
new non-methane field brown dwarfs in the first $\sim$1\% of the 2MASS
survey. They did not find any clear examples of methane-absorbing
brown dwarfs. Using the sensitivity, field coverage and other
particulars of 2MASS in Equation~3 yields $\langle N\rangle=0.7$ for
the expected number of methane absorbing brown dwarfs, consistent with
their finding none. Setting $f_{CH_4}\equiv$1 in equation~3 gives
$\langle N\rangle=4.2$ for the {\it total} number of brown dwarfs
expected, not just those with methane absorption.  The 90\% confidence
interval of a distribution with mean 4.2 marginally includes an
experiment with 7 detections. DENIS has qualitatively lower K~band
sensitivity, and it is no surprise that no methane dwarfs appeared in
the Delfosse \etal. preliminary survey.

We can also compare our calculations to the recent report
(\markcite{Bu_99}Burgasser \etal. 1999) of 4 methane-absorbing brown
dwarfs in a subsequent 2MASS survey area considerably larger than that
presented in \markcite{Ki_99}Kirkpatrick \etal. (1999). Scaling the
expectations for methane brown dwarfs in the Kirkpatrick \etal. sample
up to the larger area gives $\langle N\rangle\approx 3.5$, completely
consistent with the number of objects discovered.

These calculations highlight one of the strengths of using the methane
absorbing brown dwarfs to constrain the substellar mass function. The
survey volume for non-CH$_4$ absorbers is overwhelmingly dominated by
the youngest, hottest, objects, and is therefore very dependent on the
selection of the lower age boundary $t_{min}$ in Equation~3.  However,
the absence of methane in the atmospheres of such hot objects makes the
determination of $\langle N\rangle$ for CH$_4$-absorbers insensitive
to assumptions about $t_{min}$.  For example, setting the lower
integration limit to 0.1~Gyr more than doubles the expected {\it
total} number of substellar objects, but increases $\langle N\rangle$
for the methane dwarfs by less than 10\%.  (Note also that reducing
$t_{min}$ to 0.3~Gyr brings the prediction of Equation~3 into complete
agreement with the $N=7$ total brown dwarfs found by 2MASS.)

Two important caveats deserve mention. First, our identification
technique depends on the presence of detectable methane.  Aside from
the temperature effects discussed above, there are theoretical
calculations which suggest that cloud formation in the stellar
atmosphere may suppress the contrast in the methane band features for
certain combinations of effective temperature and surface gravity
(see, for example, \markcite{Ma_99}Marley \etal. 1999). Based on our
measurements of Gl~229b and the spectra of \markcite{St_99}Strauss
\etal. 1999 and \markcite{Bu_99}Burgasser \etal. 1999, we would have
easily identified all of the half-dozen known objects in this
class. Nevertheless, any survey, including ours, which depends on a
partial darkening of the K band may be biased, should cloud formation
prove to be an important process for these objects.

The second caveat concerns stellar multiplicity. Gl~229b is in a
binary system only 6~pc from the sun. We were easily able to detect
and identify the unusual character of Gl~229b using our filter set
(Figure~1), but it would have been impossible to separate the stars at
any distance approaching the $r_{max}\sim 100$~pc cited above.  The
problem of dynamic range and angular resolution is also a central
issue for the all-sky surveys, since they typically have spatial
sampling 3-5 times coarser than ours. (None of the recently reported
methane dwarfs are in obvious multiple systems.)  And for everyone, it
is an uncomfortable fact of life that the majority of the search
volume lies at the greatest distances for which the spatial resolution
is poorest.

\acknowledgements

The authors are grateful to the High-Z Supernova Search Team, and
particularly Nick Suntzeff and Ricardo Covarrubias, for making the
R~band observations available for this survey. We are also thankful
for the constructive comments of an anonymous referee. TMH
acknowledges fruitful discussions with Coryn Bailer-Jones, David
Barrado y Navascues, Christoph Leinert, Reinhard Mundt, Michael Meyer,
and Massimo Robberto.

\newpage

%
%
%

\figcaption[]{
(top) Profiles of our custom filters superimposed on a spectrum of
Gl~229b.  Tailoring the filter profiles to the expected methane
feature produces an efficiency gain of a factor 12 over 1\% narrow
band filters (\eg. Rosenthal \etal. 1996).  (bottom) Images of Gl~229b
in the continuum K$_C$ (left) and absorption K$_A$ (right) filters. 
The primary star image is saturated in the gray regions
}

\figcaption[]{
(top) J~band (left) and R~band (right) images of a candidate with
(R-J)$>$7.1.  (bottom) $K_C - K_A$ Colour vs Stellarity for the 44
objects with extreme (R-J) colour.  Also plotted are the measurements
for Gl~229b. None of the candidates resembles a methane-absorbing
brown dwarf.  }

\end {document}
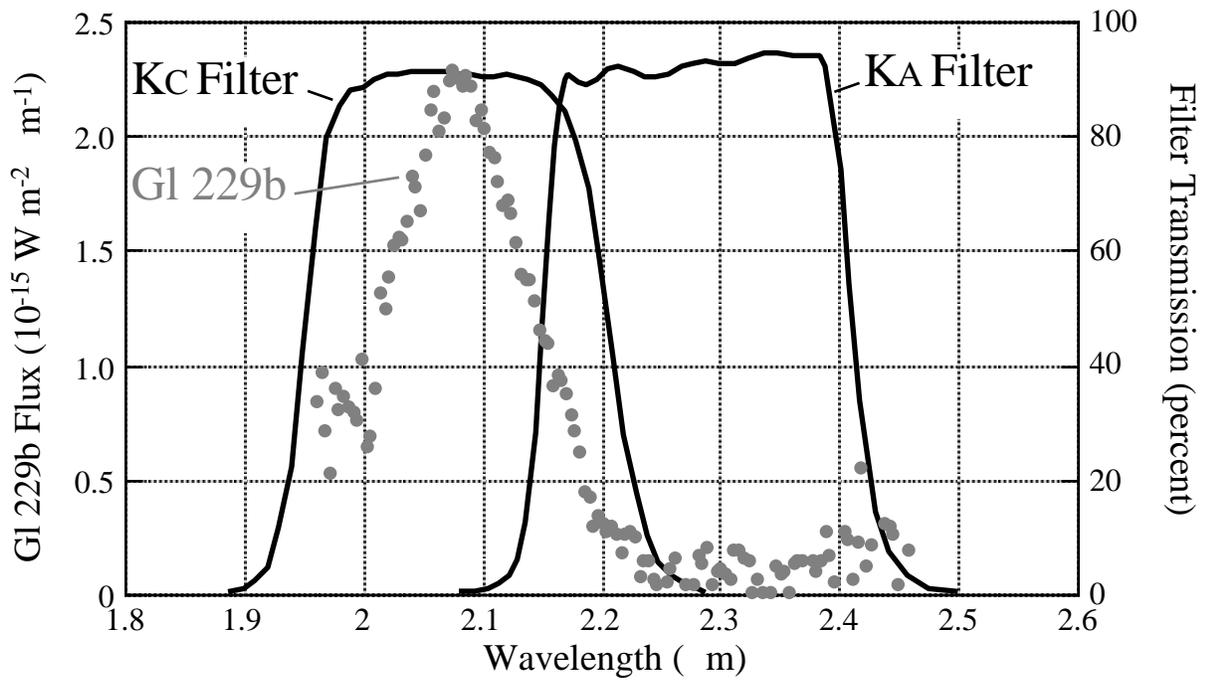
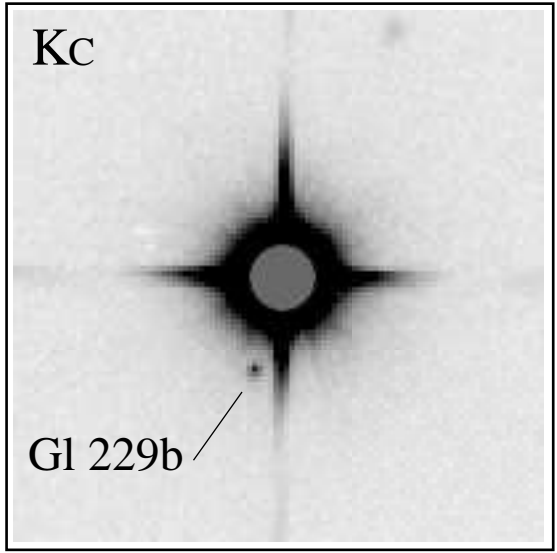
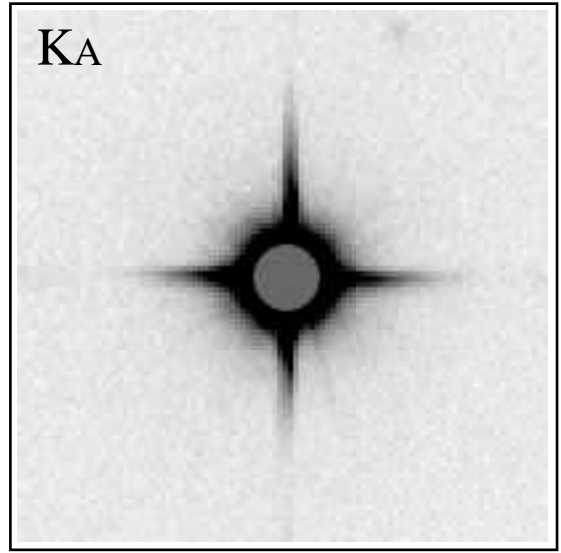
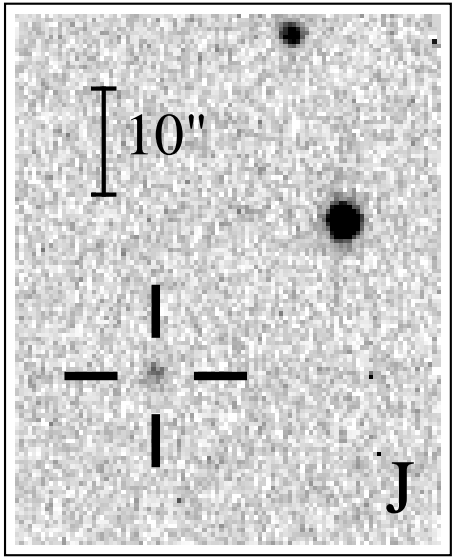
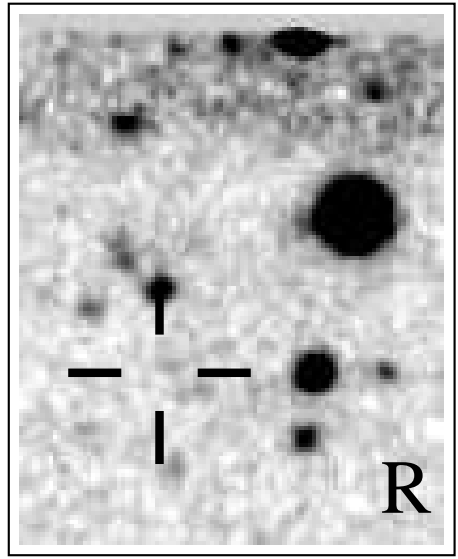
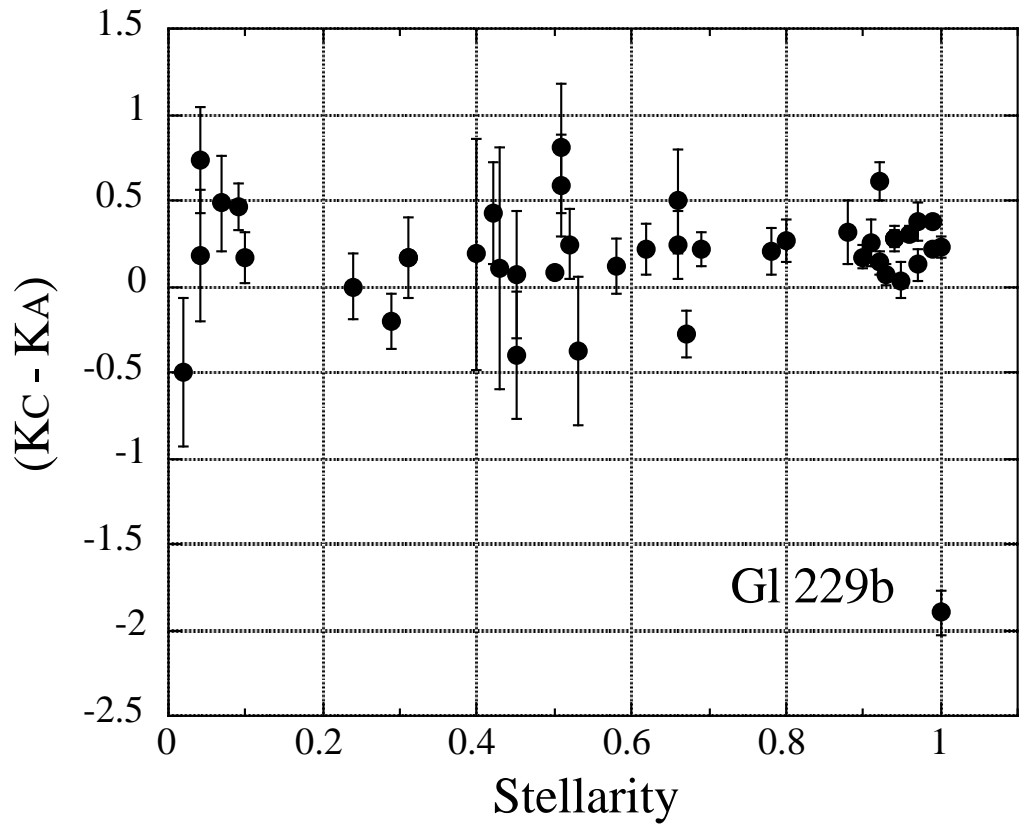
